\documentclass[twocolumn]{aastex63} 

\usepackage{graphicx} 
\usepackage{float} 
\usepackage{wrapfig} 
\usepackage{lipsum} 
\usepackage{hyperref}
\usepackage{times}
\usepackage{amsmath}
\usepackage{graphicx}
\usepackage{subfigure}
\usepackage{placeins}
\usepackage{hyperref}
\usepackage{gensymb}
\usepackage{upgreek}
\usepackage{natbib}

\usepackage{graphicx}
\usepackage{subfigure}
\usepackage{multirow}
\usepackage{comment}
\usepackage{natbib}
\usepackage{hyperref}
\usepackage{mathtools}
\usepackage{mathrsfs}
\usepackage{fontenc}
\usepackage{color}
\usepackage{url}
\usepackage{hyperref}
\usepackage{gensymb}
\usepackage{pifont}
\graphicspath{{./}}

\newcommand{\eg}{\textrm{e.g., }}

\newcommand       \Angstrom     {\,{\rm \AA}}

\newcommand       \K            {\,{\rm K}}
\newcommand       \mum          {{\rm \mu m}}

\shorttitle{Spatially Resolved PAH Emission in M51}
\shortauthors{Zhang, Ho \& Xie}

\begin{document}

\title{A Method to Extract Spatially Resolved Polycyclic Aromatic Hydrocarbon Emission from Spitzer Spectra:  Application to M51}

\author[0000-0003-4937-9077]{Lulu Zhang}
\affiliation{Kavli Institute for Astronomy and Astrophysics, Peking University, Beijing 100871, China}
\affiliation{Department of Astronomy, School of Physics, Peking University, Beijing 100871, China}

\author[0000-0001-6947-5846]{Luis C. Ho}
\affiliation{Kavli Institute for Astronomy and Astrophysics, Peking University, Beijing 100871, China}
\affiliation{Department of Astronomy, School of Physics, Peking University, Beijing 100871, China}

\author[0000-0002-9707-1037]{Yanxia Xie}
\affiliation{Kavli Institute for Astronomy and Astrophysics, Peking University, Beijing 100871, China}

\email{l.l.zhang@pku.edu.cn}

\begin{abstract}

The mid-infrared spectrum contains rich diagnostics to probe the physical properties of galaxies, among which the pervasive emission features from polycyclic aromatic hydrocarbons (PAHs) offer a promising means of estimating the star formation rate (SFR) relatively immune from dust extinction. This paper investigates the effectiveness of PAH emission as a SFR indicator on sub-kpc scales by studying the Spitzer/IRS mapping-mode observations of the nearby grand-design spiral galaxy M51. We present a new approach of analyzing the spatial elements of the spectral datacube that simultaneously maximizes spatial resolution and spatial coverage, while yielding reliable measurements of the total, integrated $5-20\,\mu$m PAH emission. We devise a strategy of extracting robust PAH emission using spectra with only partial spectral coverage, complementing missing spectral regions with properly combined mid-infrared photometry. We find that in M51 the PAH emission correlates tightly with the extinction-corrected far-ultraviolet, near-ultraviolet, and H$\alpha$ emission, from scales~$\sim~0.4$ kpc close to the nucleus to 6~kpc out in the disk of the galaxy, indicating that PAH serves as an excellent tracer of SFR over a wide range of galactic environments. But regional differences exist. Close to the active nucleus of M51 the 6.2~$\mum$ feature is weaker, and the overall level of PAH emission is suppressed. The spiral arms and the central star-forming region of the galaxy emit stronger 7.7 and 8.6~$\mu$m PAH features than the inter-arm regions. 
\end{abstract}

\keywords{dust, extinction --- galaxies: individual (M51) --- galaxies: ISM --- galaxies: star formation --- infrared: ISM}

\section{Introduction} 

Much effort has been devoted to calibrating empirical estimators of star formation activity in galaxies (\citealt{Kennicutt & Evans 2012}, and references therein), as accurate measurements of star formation rate (SFR) are vital to understand many aspects of galaxy evolution. The mid-infrared (MIR) spectral region furnishes abundant diagnostics on the physical conditions of dust and gas, which can be used to probe the sources of heating and excitation of the interstellar medium (ISM). For example, the high-ionization fine-structure lines of [Ne~{\small V}]~14.3 and 24.3~$\mum$ and [O~{\small IV}]~25.9~$\mum$ are indicative of a hard radiation field, such as that typically emitted by highly accreting active galactic nuclei (AGNs; e.g., \citealt{Armus et al. 2007}; \citealt{Melendez et al. 2008}), while the lower-ionization species [Ne~{\small II}]~12.8~$\mum$ and [Ne~{\small III}]~15.6~$\mum$ can effectively trace star-forming regions in normal (\citealt{Ho & Keto 2007}) and active (\citealt{Zhuang et al. 2019}) galaxies. Apart from these and other narrow ionic lines, the bulk of the spectral window from $\sim 5$ to 20~$\mu$m is dominated by a series of prominent, broad emission features identified with polycyclic aromatic hydrocarbons (PAHs). Photoelectrically heated by single ultraviolet (UV) photons (\citealt{Allamandola et al. 1989}; \citealt{Bakes & Tielens 1994}; \citealt{Li & Draine 2002}), PAH emission is widely regarded as an effective tracer of far-UV (FUV) flux and hence recent star formation (\citealt{Peeters et al. 2004}), if PAHs are mostly associated with photo-dissociation regions surrounding H~{\small II} regions (\citealt{Helou et al. 2004}; \citealt{Rho et al. 2006}; \citealt{Smith & Brooks 2007}; \citealt{Tielens 2008}) and not with non-ionizing stars (\citealt{Crocker et al. 2013}; \citealt{Lu et al. 2014}). Extensive observations with the Infrared Space Observatory (\citealt{Kessler et al. 1996}), the Infrared Spectrograph (IRS; \citealt{Houck et al. 2004}) on the Spitzer Space Telescope (\citealt{Werner et al. 2004}), and the Infrared Camera (IRC) on the AKARI satellite (\citealt{Onaka et al. 2007}) have revealed that PAHs are prevalent in many astronomical environments (\eg \citealt{Genzel et al. 1998}; \citealt{Armus et al. 2007}; \citealt{Galliano et al. 2008}; \citealt{Kaneda et al. 2012}; \citealt{Yamada et al. 2013}; \citealt{Cortzen et al. 2019}). Contributing as much as $10\%-20\%$ of the total IR emission in star-forming galaxies (\citealt{Smith et al. 2007b};  \citealt{Diamond-Stanic & Rieke 2010}; \citealt{Xie et al. 2018a}), PAH emission has the potential to serve as an effective SFR estimator for galaxies at low and high redshifts, especially with the advent of upcoming facilities such as JWST and SPICA. PAH emission has the additional advantage of being much less sensitive to dust extinction than SFR tracers at shorter wavelengths.

There have been numerous attempts to calibrate PAH emission against other traditionally well-established SFR indicators. One set of studies employs broadband observations using filters that capture strong PAH features, such as the 7.7~$\mum$ feature using the Spitzer IRAC4 8~$\mum$ band (e.g., \citealt{Calzetti et al. 2005, Calzetti et al. 2007}; \citealt{Wu et al. 2005}; \citealt{Zhu et al. 2008}; \citealt{Battisti et al. 2015};  \citealt{Mahajan et al. 2019}) and the 11.3~$\mum$ feature using the WISE W3 12~$\mum$ band (e.g., \citealt{Donoso et al. 2012}; \citealt{Lee et al. 2013}; \citealt{Cluver et al. 2017}). While this approach has the merit that it can be applied to large samples of objects, the accuracy of broadband calibrations is affected by complications from continuum contamination and, in the case of the WISE W3 band, 9.7~$\mum$ silicate absorption (e.g., \citealt{Donoso et al. 2012}; \citealt{Cluver et al. 2014}). Where spectroscopic observations are available, individual PAH features can be decomposed (\citealt{Peeters et al. 2002}; \citealt{Marshall et al. 2007}; \citealt{Smith et al. 2007b}; \citealt{Xie et al. 2018a}), and SFR calibrations have been devised based on individual PAH bands (6.2, 7.7, and 11.3~$\mum$), combinations thereof, or integrated across all features over a broad spectral range, for galaxies near (e.g., \citealt{Farrah et al. 2007}; \citealt{Shi et al. 2007}; \citealt{Sargsyan & Weedman 2009}; \citealt{Treyer et al. 2010}; \citealt{Diamond-Stanic & Rieke 2012}; \citealt{Shipley et al. 2016}; \citealt{Xie & Ho 2019}) and far (e.g., \citealt{Pope et al. 2008}; \citealt{Menendez-Delmestre et al. 2009}).  

When spatially averaged over large scales, PAH emission is not only pervasive but also remarkably invariant in terms of its overall spectral characteristics across a wide range of galaxy environments (\citealt{Xie et al. 2018a}). This gross similarity belies the spectral diversity that surfaces upon closer scrutiny, both within individual galaxies and among different galaxies. For instance, while the 6.2, 7.7, and 11.3~$\mum$ bands are closely coupled, the ratios between these bands vary depending on the excitation mechanisms of PAHs and the radiation field to which PAHs are exposed (e.g., \citealt{Pagani et al. 1999}; \citealt{Galliano et al. 2008}; \citealt{Gordon et al. 2008}; \citealt{Calzetti 2011}; \citealt{Maragkoudakis et al. 2018}). Metallicity affects the hardness of the radiation field and hence the manner in which PAHs are processed (\citealt{Smith et al. 2007b}), motivating certain investigators to include metallicity explicitly into their PAH-based SFR calibrations (e.g., \citealt{Shipley et al. 2016}; \citealt{Xie & Ho 2019}). The contribution by non-ionizing stars to the FUV flux poses another complication (\citealt{Peeters et al. 2004}). In their spatially resolved analysis of M51 based on Spitzer imaging data, \cite{Calzetti et al. 2005} note that the 8~$\mum$ emission (dominated by the 7.7~$\mum$ PAH feature) is not directly proportional to the number of ionizing photons, likely as a result of the mixture of UV radiation from current and recent star formation. Since the early work of \cite{Aitken & Roche 1985}, it has been recognized that the harsh radiation field in the vicinity of an AGN may be particularly hostile to the survival of PAHs (\citealt{Voit 1992}), leading to strong variations in the strength of the 5--8~$\mum$ PAH features (\citealt{Smith et al. 2007b}; \citealt{Diamond-Stanic & Rieke 2010}).

Spatially resolved studies are critically important to help disentangle these and other factors that contribute to the substantial systematic uncertainties that impact SFR estimates based on the PAH features. Better insights will also emerge into the fundamental physics of the ISM. In this first of a series of works, we utilize Spitzer IRS mapping-mode observations to develop an optimal method to extract spatially resolved, sub-kpc PAH measurements of nearby galaxies. Following \cite{Xie et al. 2018a}, our emphasis is on securing a robust flux of the total integrated emission over all of the PAH features in the spectral region 5--20~$\mu$m, simultaneously maximizing spatial coverage and spatial resolution. We demonstrate, by combining the IRS spectra with suitably chosen MIR imaging data, that we can derive reliable PAH fluxes even without full spectral coverage. As a pilot study, the current paper focuses on the nearby face-on, grand-design spiral galaxy M51, which, as part of the Spitzer Infrared Nearby Galaxies Survey (SINGS; \citealt{Kennicutt et al. 2003}), enjoys a plethora of ancillary multiwavelength data that can be utilized to our advantage. In addition to forming stars copiously (\citealt{Calzetti et al. 2005}), M51 hosts a radio-quiet (\citealt{Ho & Ulvestad 2001}) Seyfert 2 nucleus (\citealt{Ho et al. 1997}), affording the opportunity to investigate the influence of its AGN on PAH emission. At a distance of 8.2 Mpc (\citealt{Kennicutt et al. 2003}), 1\arcsec\ corresponds to 40~pc.

This paper is structured as follows. Section~\ref{section:sec2} describes the observational material and procedures for data processing. Section~\ref{section:sec3} presents the methodology to measure PAH emission, both for the case using complete IRS low-resolution spectra and when only partial spectral coverage is available. Section~\ref{section:sec4} discusses some simple scientific applications, including spatially resolved correlations between PAH emission and other star formation indicators, as well as the variation of PAH emission with environment. A summary of the results is given in Section~\ref{section:sec5}.

\section{Observational Material}\label{section:sec2}

\subsection{Datasets}
 
We use Spitzer/IRS spectral mapping-mode observations to study spatially resolved PAH emission in nearby galaxies. In this first application, we focus on M51 to develop and illustrate our technique. While the general framework of our method performs spectral decomposition over the full $5-38\,\mum$ bandpass of IRS, not all galaxies in the IRS archive have complete spectral coverage. This motivates us to extend our fitting method to the regime in which only partial IRS observations are available, supplementing the missing spectral coverage with properly matched photometric data points adapted from MIR images (Section~\ref{section:sec3.2}). Broadband UV and narrowband H$\alpha$ images are also used to examine the spatial correspondence between these traditional star formation tracers and PAH emission. Table~\ref{tab:table1} summarizes the multiwavelength photometric datasets used for analyzing M51.

\begin{deluxetable}{cccccc}[!ht]
\tabletypesize{\small}
\tablecolumns{6}
\tablecaption{Multiwavelength Photometric Datasets for M51}
\tablehead{
\colhead{Telescope} & \colhead{Filter} & \colhead{$\lambda_{\rm eff}$} & \colhead{FWHM} & \colhead{$\sigma_{\rm cal}$} & \colhead{References}\\
\colhead{} & \colhead{} & \colhead{$(\mu m)$} & \colhead{(\arcsec) } & \colhead{(\%)} & \colhead{}
}
\startdata
GALEX & FUV & 0.153 & 4.2 & 5 & 1, 2\\
GALEX & NUV & 0.231 & 5.3 & 3 & 1, 2\\
\hline
{KPNO}     & H$\alpha$ & 0.657 & 1.9 & 15 & 3, 4\\
\hline
Spitzer & IRAC4 & 7.87 & 2.0 & 10 & 3, 5\\
Spitzer & MIPS24 & 23.68 & 6.0 & 4 & 3, 6\\
\hline
AKARI & IRC S7  & 7.0 & 2.9 & 2.3 & 7, 8\\
AKARI & IRC S11 & 11.0 & 3.3 & 2.4 & 7, 8\\
AKARI & IRC L15 & 15.0 & 4.6 & 2.8 & 7, 8\\
AKARI & IRC L24 & 24.0 & 6.7 & 4.7 & 7, 8\\
\hline
WISE &  W3 & 11.56 & 6.5 & 4.5 & 9,10\\
\enddata
\tablecomments{Col. (1): Telescope. Col. (2): Filter. Col. (3): Effective wavelength of filter. Col. (4): FWHM of the PSF. Col. (5): Calibration uncertainty. Col. (6):  References.}
\tablerefs{(1) \citealt{Morrissey et al. 2007}; (2) \citealt{Gil de Paz et al. 2007}; (3) \citealt{Calzetti et al. 2005}; (4) \citealt{Dale et al. 2007}; (5) \citealt{Fazio et al. 2004}; (6) \citealt{Engelbracht et al. 2007}; (7) \citealt{Arimatsu et al. 2011}; (8) \citealt{Egusa et al. 2013}; (9) \citealt{Wright et al. 2010}; (10) \citealt{Jarrett et al. 2013}.}
\label{tab:table1}
\end{deluxetable}

\subsubsection{Spitzer/IRS Mapping-mode Spectroscopy}

The IRS has both a low-resolution and a high-resolution module, each with a short slit and a long slit to capture a different wavelength range. This paper only uses the low-resolution spectra, which themselves encompass two modes, each further split into two spectral orders: (1) the short-low (SL) mode covers $5.2-7.7\ \mu$m in the short-low 2 (SL2) order with a 3\farcs6$\times$57\arcsec\ slit and $7.4-14.5\ \mu$m in the short-low 1 (SL1) order with a 3\farcs7$\times$57\arcsec\ slit; (2) the long-low (LL) mode covers $14.0-21.3\ \mu$m in the long-low 2 (LL2) order with a 10\farcs5$\times$168\arcsec\ slit and $19.5-38.0\ \mu$m in the long-low 1 (LL1) order with a 10\farcs7$\times$168\arcsec\ slit. The full-width at half maximum (FWHM) resolution of the low-resolution spectra varies from $\lambda/\Delta\lambda\approx$ 64 to 128 in each mode from one end of the spectrum to the other. The mapping-mode observations were conducted by scanning the slit to cover large sections of the galaxy.

\begin{figure}[!ht]
\center{\includegraphics[width=1\linewidth]{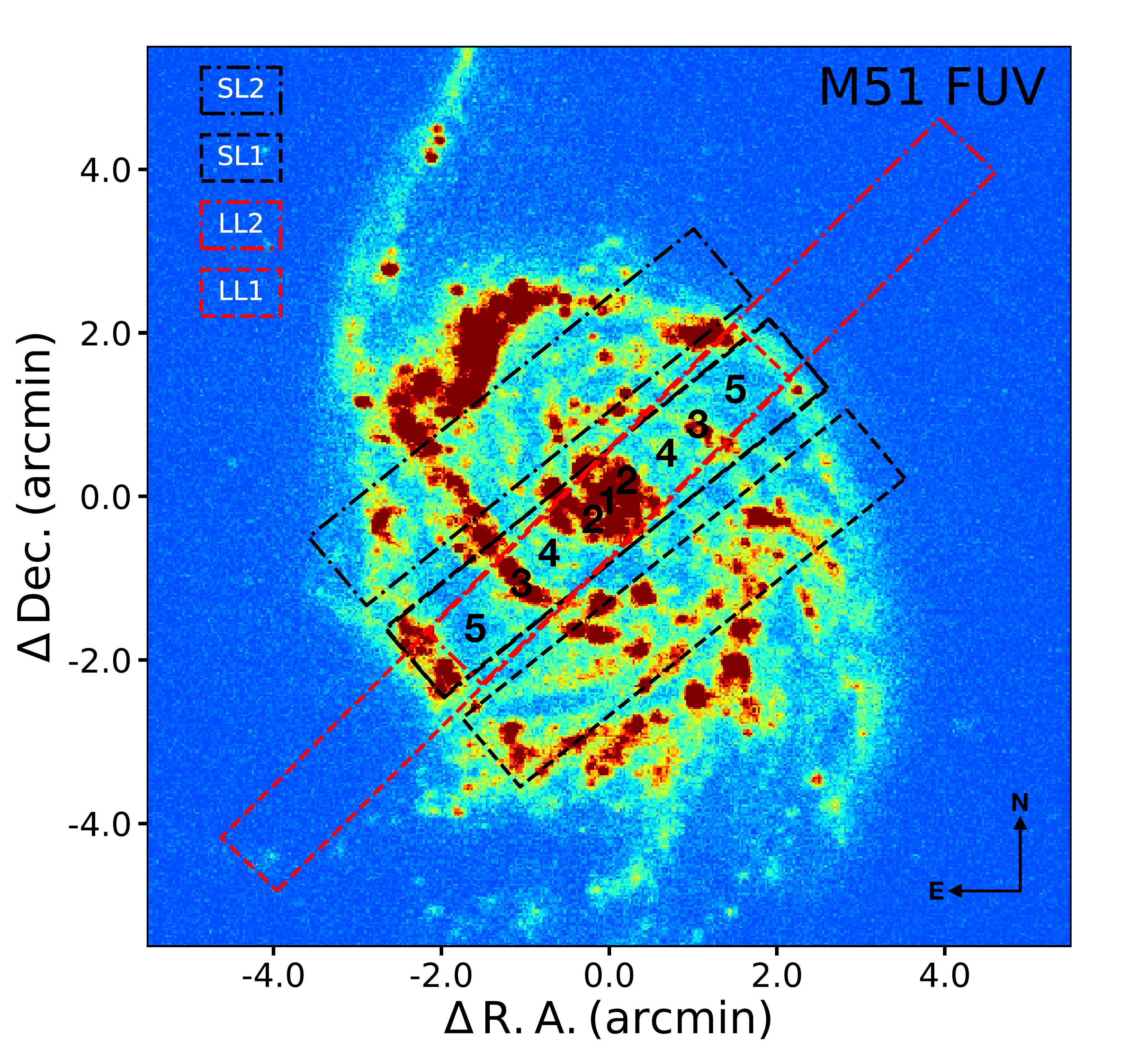}}
\caption{{Setup for the IRS spectral mapping-mode observations of M51, showing the areas covered by the SL and LL slits overlaid on the GALEX FUV image of the galaxy. The five galactic regions discussed in Section~\ref{section:sec4} are numbered schematically.} \label{fig:fig01}}
\end{figure}

We begin with the basic calibrated data of the low-resolution mapping-mode observations of M51 (program 20138; PI: K. Sheth) stored in the {\rm Spitzer}\ Heritage Archive\footnote{\url{https://sha.ipac.caltech.edu/applications/Spitzer/SHA/}}. We further process these basic calibrated data products using {\tt CUBISM} (\citealt{Smith et al. 2007a}) to combine them into a three-dimensional spectral datacube. Background subtraction for the SL and LL spectra was achieved through dedicated off-source and outrigger observations. Figure~\ref{fig:fig01} illustrates the spatial coverage of the mapping-mode observations of M51. We concentrate on the region bisecting the center of the galaxy that contains complete low-resolution spectral coverage, namely the area where the SL and LL modes overlap. For the purposes of subsequent discussion, we divide all the pixels in the area of interest roughly into five categories: (1) the ``nuclear region'', which includes the AGN and its immediately surrounding pixels; (2) the ``central region'', which covers the central $\sim 2$ kpc-diameter star-forming region (\citealt{Scoville et al. 2001}; \citealt{Lamers et al. 2002}), except for the pixels included in the nuclear region; (3) ``spiral arms'', which includes the pixels of the spiral arms located on diametrically opposite sides of the galaxy roughly $\sim 3$ kpc from the center; (4) ``inner ISM'', which covers the regions $\sim 2$ kpc from the nucleus located in between the central region and spiral arms; and (5) ``outer ISM'', which samples the inter-arm region $\sim 4$ kpc from the nucleus exterior to the spiral arms.

\subsubsection{Imaging}\label{section:sec2.1.2}

Section~\ref{section:sec3.2} describes our method of extracting PAH measurements by combining incomplete IRS spectra with appropriately chosen MIR photometric data that fill in the missing spectral coverage. As summarized in Table~\ref{tab:table1}, M51 has wide-field imaging observations from the Spitzer/IRAC (3.6, 4.5, 5.8, 7.9~$\mu$m) and Spitzer/MIPS (24, 70, 160~$\mu$m) channels from SINGS\footnote{\url{https://irsa.ipac.caltech.edu/data/SPITZER/SINGS/galaxies/ngc5194.html}}. We complement these with additional imaging from WISE\footnote{\url{https://irsa.ipac.caltech.edu/applications/wise/}} (\citealt{Wright et al. 2010}) and AKARI/IRC\footnote{\url{https://www.ir.isas.jaxa.jp/AKARI/Archive/Images/IRC_Images/search/}} (\citealt{Egusa et al. 2016}). The AKARI/IRC MIR-L frames of M51 are contaminated by Earth shine light, which we remove following \cite{Egusa et al. 2013}. We adjust aperture correction differences stemming from differences in the photometric calibration between point sources and extended sources (\citealt{Dale et al. 2007}; \citealt{Engelbracht et al. 2007}; \citealt{Arimatsu et al. 2011}).

We utilize the far-ultraviolet (FUV; 1350 \AA--1750 \AA) and near-ultraviolet (NUV; 1750 \AA--2750 \AA) images from the GALEX Ultraviolet Atlas of Nearby Galaxies\footnote{\url{https://archive.stsci.edu/prepds/galex_atlas/}} (\citealt{Gil de Paz et al. 2007}). The continuum-subtracted narrowband H$\alpha$ image of M51 comes from the ancillary data of SINGS. The bandpass of the H$\alpha$ filter includes [N~{\small II}]~$\lambda\lambda6548, 6584$, and we correct for this contamination following the procedure of \cite{Kennicutt et al. 2008}.

\begin{figure*}[!ht]
\center{\includegraphics[width=1\textwidth]{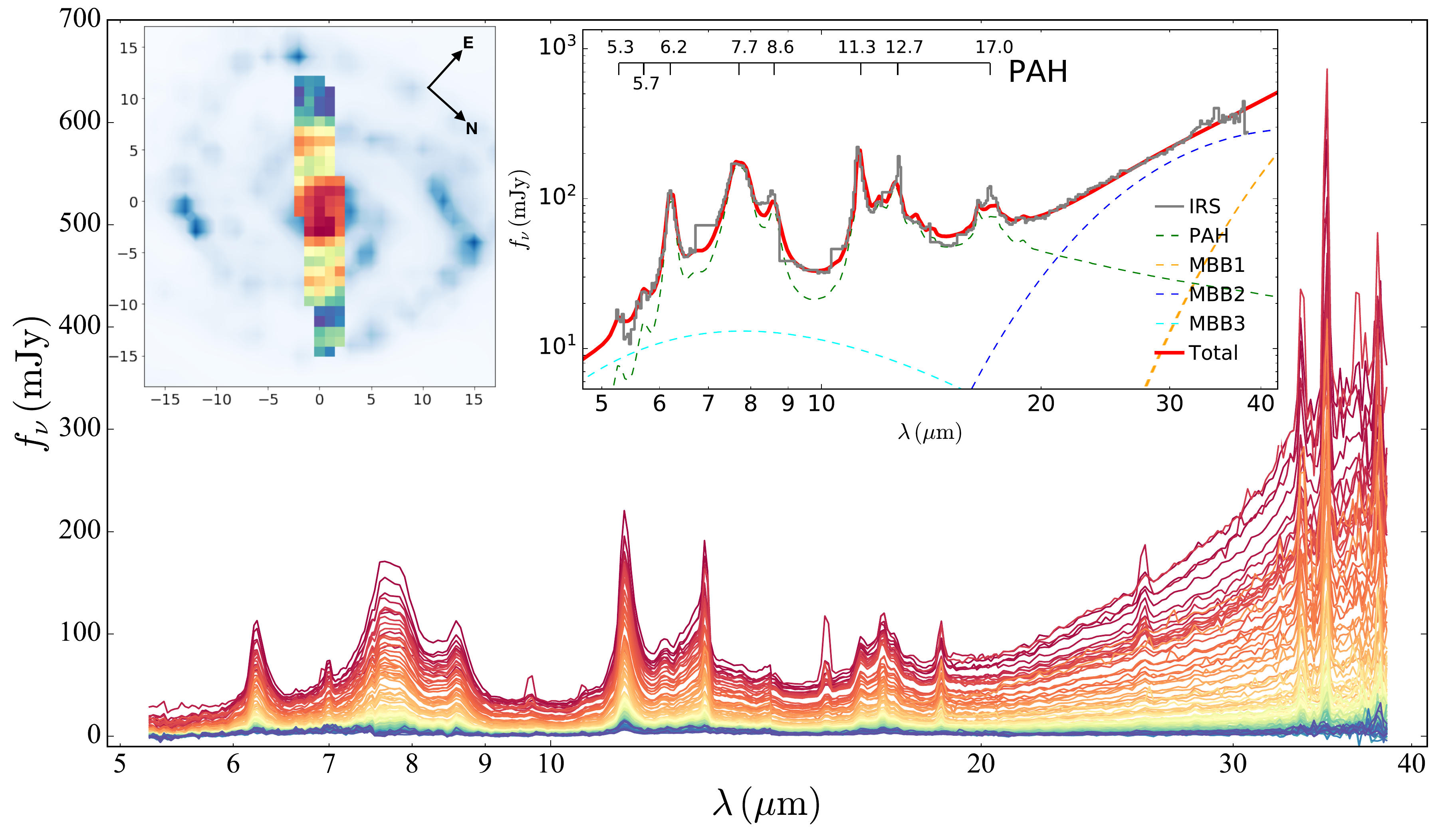}}
\caption{Spectra extracted in different spatial bins that have complete IRS low-resolution spectra, color-coded (from blue to red) according to the integrated flux from $5-20\,\mu$m. The spatial distribution of the bins are shown in the image in the upper-left corner. The top-right inset illustrates the PAH decomposition methodology for an example spectrum (black solid line) following the method described in Section~\ref{section:sec3.1}. The model (red solid line) is composed of a PAH template and three continuum components, each represented by a modified black body (MBB). Prominent PAH emission features are marked.
\label{fig:fig02}}
\end{figure*}

\subsection{Data Processing}

\subsubsection{Background Subtraction}

Background subtraction for the spectral cubes was performed as part of the data processing using {\tt CUBISM}, using the dedicated off-source background and outrigger observations. The median contribution of the background is $\sim1\%$, increasing from the nuclear (central $\sim 400$ pc) region to the outer diffuse region ($\sim 4$ kpc from the nucleus), with a maximum of $\sim5\%$. Although the publicly released images are already background-subtracted\footnote{\url{https://irsa.ipac.caltech.edu/data/SPITZER/SINGS/doc/sings_fifth_delivery_v2.pdf}}, we find that some images display evident background residuals that need further treatment. We remove background residuals by fitting and subtracting a two-dimensional polynomial function after masking out real sources.

\subsubsection{Convolution}

The IRS spectral datacube has a point-spread function (PSF) that increases monotonically with increasing wavelength, from ${\rm FWHM} \approx 1\farcs5-3\farcs5$ for the SL mode to $\sim 4\farcs5-9\farcs0$ for the LL mode. By comparison, the various imaging datasets used in this work have PSF sizes with ${\rm FWHM} \approx 2\farcs0-6\farcs7$ (Table~\ref{tab:table1}). To jointly analyze the spectroscopic and photometric data, we need to convolve the two-dimensional images and three-dimensional spectral datacube to a common PSF to ensure that they cover the same physical scale. To guarantee the accuracy of the convolution kernels, we choose a common FWHM of $10\farcs0$, which is slightly broader than the largest common PSF (${\rm FWHM} = 9\farcs0$) among all the images and slices of the spectral datacube. \cite{Aniano et al. 2011} provide convolution kernels, described by Moffat functions and a sum of Gaussians\footnote{\url{https://www.astro.princeton.edu/~ganiano/Kernels/Ker_2012/Kernels_fits_Files/Low_Resolution/}}, for the various cameras on Spitzer, GALEX, WISE, and ground-based optical telescopes. We assume a Moffat function for the PSF of the H$\alpha$ image. For AKARI, the convolution kernels are constructed according to the procedures in \cite{Aniano et al. 2011} based on the PSFs of IRC images.
 
The situation for the IRS spectral datacube is complicated because the PSF varies with wavelength, and the theoretical PSF is not well explored. To perform a convolution on the spectral cube, we first create a theoretical PSF for each wavelength using {\tt sTinyTim} (\citealt{Krist 1995}), a widely used (e.g., \citealt{Engelbracht et al. 2007}; \citealt{Smith et al. 2009}; \citealt{Bendo et al. 2010}; \citealt{Aniano et al. 2011}; \citealt{Beirao et al. 2012}) program to generate theoretical PSF models specifically for Spitzer instruments\footnote{\url{https://irsa.ipac.caltech.edu/data/SPITZER/docs/dataanalysistools/tools/contributed/general/stinytim/}}. We then follow \cite{Aniano et al. 2011} to construct the convolution kernels using the {\tt create\_matching\_kernel} function in the {\tt Python} package {\tt photutils}\footnote{\url{https://media.readthedocs.org/pdf/photutils/stable/photutils.pdf}}. The {\tt convolve\_fft} function in {\tt astropy.convolution} (\citealt{Astropy Collaboration et al. 2013}) is used to carry out the convolution.

Several works have tried to explore the PSF of the IRS spectral datacube (\eg \citealt{Sandstrom et al. 2009}; \citealt{Lebouteiller et al. 2010, Lebouteiller et al. 2015}; \citealt{Pereira-Santaella et al. 2010}). Among them, \cite{Lebouteiller et al. 2010} created super-sampled PSFs from the low-resolution (\citealt{Lebouteiller et al. 2010}) and high-resolution (\citealt{Lebouteiller et al. 2015}) data. Direct comparison of spectra extracted from a datacube convolved by our convolution kernel with spectra derived from a datacube convolved by the convolution kernel of \cite{Lebouteiller et al. 2010} reveals reasonably good agreement. The ratio of flux densities of the two sets of spectra has a median value and standard deviation of $1.22 \pm 0.20$. This systematic difference likely arises from the different normalization strategies adopted in the convolution kernels (V. Lebouteiller 2019, private communications). In their spatially resolved study of local infrared-luminous galaxies, \cite{Pereira-Santaella et al. 2010} used mapping-mode observations of calibration stars in the Spitzer archive to characterize the PSF of their IRS spectral datacubes. The PSFs provided by \cite{Pereira-Santaella et al. 2010} are broader than those estimated by {\tt sTinyTim}. In principle, convolution kernels calculated based on wider PSFs will be less dispersed, such that the convolution of the IRS datacube would be less influenced by ``edge effects,'' as discussed later.

\subsubsection{Reprojection}
After convolving all the data to homogenize them to the same angular resolution, we reproject both the photometric images and spectral datacube into the same coordinate frame to enable pixel-based, spatially resolved analysis. We choose a final pixel size of 10\arcsec\ for the reprojected data, which corresponds to a physical scale of $\sim 0.4\, \rm kpc$ at the distance of M51. This process is realized with {\tt Montage}\footnote{\url{http://montage.ipac.caltech.edu}}, an astronomical image mosaic engine for analyzing the geometry of images on the sky, reprojecting images, rectifying background emission to a common level, and coadding images. The uncertainty of the data is reprojected as well through Gaussian propagation, using the same strategy taken by {\tt CUBISM}.

\subsubsection{Optimal Binning}\label{section:sec2.2.4}

Our desire to study spatially resolved PAH emission over the broadest range of possible environments within a given galaxy prompts us to devise a strategy for optimally binning the pixels into final spaxels that simultaneously maximizes spatial resolution (yields the largest number of spatially independent spectra) while at the same time ensuring that each spaxel has sufficient signal-to-noise ratio (S/N) to yield accurate measurement of the PAH features. Appendix~\ref{sec:appexA} describes an experiment performed to determine the fractional uncertainty on recovering the integrated $5-20\,\mu$m PAH flux as a function of S/N. Depending on the particular science goal, Figure~A can be used as a guide to choose the desired minimum S/N criterion. For a spectral datacube consisting of $N$ spaxels, the signal $\mathcal{S}_{i}$ and noise $\mathcal{N}_{i}$ of spaxel $i$ are given, respectively, by

\begin{equation}
\label{equ:equSN}
\begin{split}
\mathcal{S}_{i}     & = \frac{1}{\Delta\lambda}\int_{\Delta\lambda}f_{i}(\lambda)d\lambda \\ 
\mathcal{N}_{i}^{2} & = \frac{1}{\Delta\lambda}\int_{\Delta\lambda}\sigma_{i}^{2}(\lambda)d\lambda,
\end{split}
\end{equation}

\noindent
where $f_{i}(\lambda)$ is the flux density at a given wavelength and $\sigma_{i}(\lambda)$ is its corresponding uncertainty. Our binning procedure begins by searching for the spaxel with the most significant signal, $\mathcal{S}_{i}$. If the spaxel meets the minimum desired S/N, it is retained as a separate bin; otherwise, we bin the spaxel with one or more adjacent spaxels until their combined S/N exceeds the minimum desired threshold, prioritized according to their ``similarity'' (\citealt{Abdurro'uf & Akiyama 2017}). For spaxels $a$ and $b$, the similarity of their spectra is quantified as

\begin{equation}
\label{equ:equSIM}
\chi^{2} = \sum_{\lambda}\frac{(f_{a,\lambda}-s_{a,b}f_{b,\lambda})^{2}}{\sigma_{a,\lambda}^{2}+\sigma_{b,\lambda}^{2}},
\end{equation}

\noindent
where $f_{a,\lambda}$ and $f_{b,\lambda}$ are their flux density, $\sigma_{a,\lambda}$ and $\sigma_{b,\lambda}$ are their corresponding uncertainty, and the normalization factor to adjust the flux density difference between the two spectra  is given by

\begin{equation}
\label{equ:equnorm}
s_{\rm ab} = \frac{\sum_{\lambda}\frac{f_{a,\lambda}f_{b,\lambda}}{\sigma_{a,\lambda}^{2}+\sigma_{b,\lambda}^{2}}}{\sum_{\lambda}\frac{f_{b,\lambda}^{2}}{\sigma_{a,\lambda}^{2}+\sigma_{b,\lambda}^{2}}}\,\,\,.
\end{equation}

\noindent
A smaller value of $\chi^2$ signifies higher similarity. We consider the similarity between spectra during the binning process in order to maximize the association of spatial structures with similar physical properties. We iterate the search for the next spaxel with the most significant signal among those not binned in the last step, and repeat the process until all the spaxels are incorporated into bins that satisfy the predetermined S/N criterion. The ancillary multiwavelength images (Section~\ref{section:sec2.1.2}) are binned to the exact same pixel size as the spectral cube to enable construction of the spatially resolved MIR spectral energy distributions (Section~\ref{section:sec3.2}). 

Figure~\ref{fig:fig02} illustrates the outcome of our optimal binning process applied to the central $40\arcsec \times 240\arcsec$ of M51, which contains full spectral (SL and LL) coverage by the low-resolution module of IRS. M51 is sufficiently bright, however, that optimal binning turns out to be unnecessary for this galaxy; all of the original 10\arcsec\ pixels have S/N $\geq$ 30, such that the fractional uncertainty on the integrated PAH flux is only $\sim 1\%$ (Appendix~\ref{sec:appexA}). The spectra of individual spaxels are color-coded (from blue to red) according to the integrated flux from $5-20\,\mu$m, and their spatial distribution is shown in the image inset on the upper-left. The top-right inset illustrates the spectral decomposition of an example spaxel using the method of \citeauthor{Xie et al. 2018a}(2018a; see Section~\ref{section:sec3.1}).

\begin{figure*}[!ht]
\center{\includegraphics[width=1\textwidth]{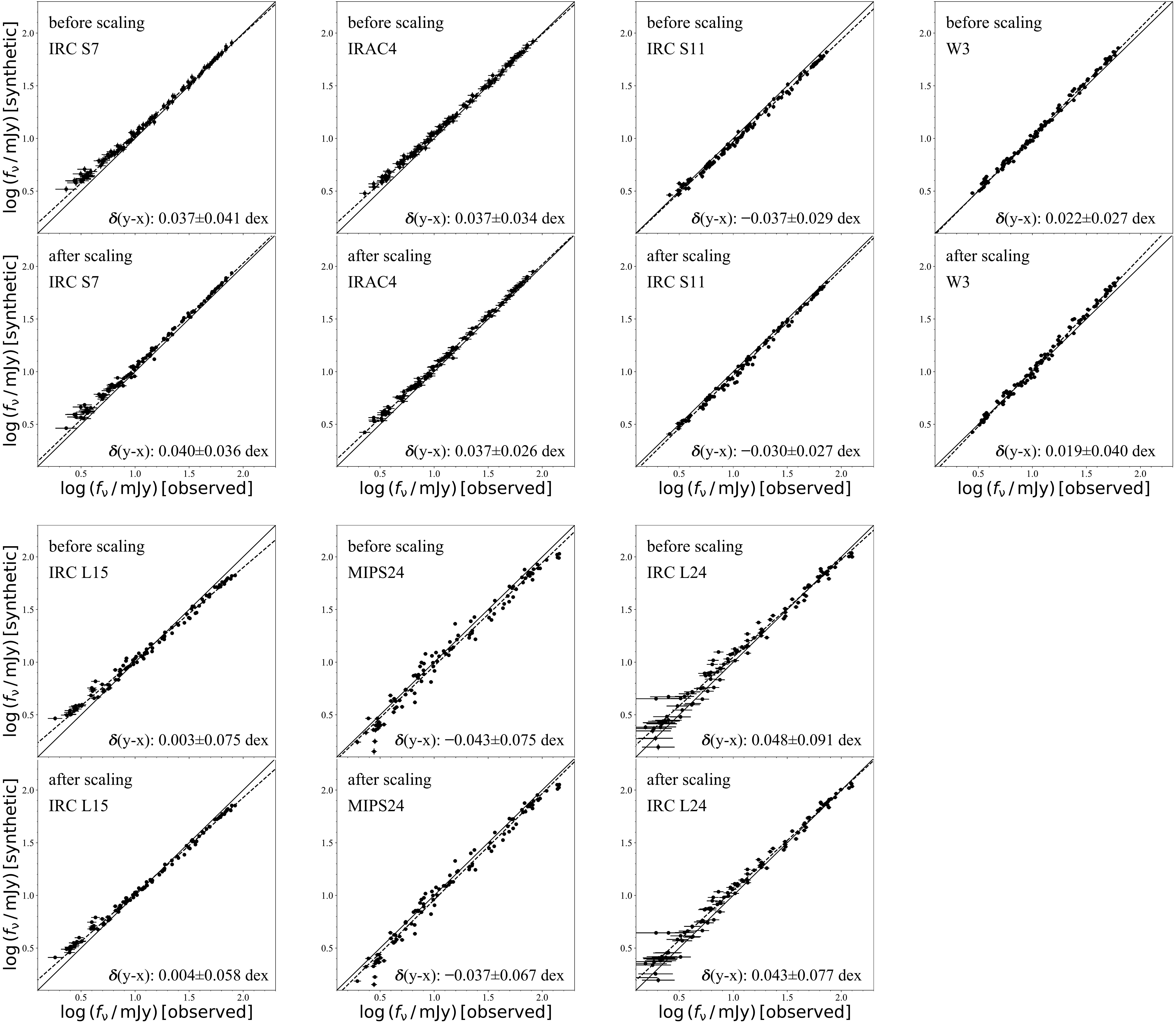}}
\caption{{Comparison between synthetic photometric flux densities and observed photometric flux densities. The solid black lines denote a 1:1 relation, while the dashed lines give the best linear fit. The median and standard deviation of the difference between the synthetic and observed flux densities are given in the lower-right corner of each panel.} \label{fig:figB}}
\end{figure*}

\subsubsection{Spectral Calibration} \label{section:sec2.2.5}

The size of the spectral cube is rectangular, with the short side smaller than the square convolution kernel. This introduces a bias to the resultant flux density.\footnote{During the convolution, regions outside of the spectral cube are set to zero flux density, which will introduce flux loss in the extracted spectra from the convolved datacube. In the opposite extreme where the size of the spectral cube is much larger than the size of the convolution kernel, only the outskirts of the spectral cube will be affected by such an effect. If the size of the spectral cube is comparable to or slightly smaller than that of the convolution kernel, the induced flux bias will be negligible because the convolution kernel values usually drop rapidly to a very low value at the outskirts.} After reprojection of the IRS datacube, we find 10 spaxels ($\sim10\%$ of the total spaxels) near the corner ($\sim 5$ kpc from the nucleus) that are influenced to varying degree by these ``edge effects.'' To mitigate against such systematics, we choose the photometric measurements to anchor the flux and scale the spectra from the binned spaxels by a factor $k$, whose optimal value is determined by minimizing $\xi$, defined as

\begin{equation}
\label{equ:equscale}
\xi^{2} = \sum_{\lambda}\left(\frac{f_{{\rm phot},\lambda} - k\, f_{{\rm IRS},\lambda}}{\sigma_{{\rm phot},\lambda}}\right)^{2},
\end{equation}
\noindent
where $f_{{\rm phot},\lambda}$ is the observed flux density of the photometric band centered at wavelength $\lambda$, $\sigma_{{\rm phot},\lambda}$ is the corresponding uncertainty, and $f_{{\rm IRS},\lambda}$ is the synthetic flux density at wavelength $\lambda$ calculated from the IRS spectrum. We use the seven MIR photometric bands listed in Table~\ref{tab:table1}, which sample the wavelength coverage of IRS. Figure~\ref{fig:figB} compares the synthetic flux densities with those observed directly from the spectra for the seven photometric bands, before and after scaling the spectra, to check whether there are systematics in our data reduction methodology. We parameterize the relation between synthetic flux densities and observed flux densities as

\begin{equation}
\label{equ:equconfirm}
f_{{\rm IRS},\lambda} = \alpha f_{{\rm phot},\lambda} + \beta,
\end{equation}
\noindent
and Table~\ref{tab:table2} summarizes the best-fit parameters before ($\alpha_{0}$, $\beta_{0}$) and after ($\alpha_{1}$, $\beta_{1}$) scaling the spectra. There is excellent consistency ($\alpha \approx 1$, $\beta \approx 0$) between the synthetic and observed flux densities for all the bands, especially after scaling the spectra, which results in smaller deviation with respect to the one-to-one relation, as indicated by the statistics given on the bottom of each panel in Figure~\ref{fig:figB}.

\begin{deluxetable}{crrrr}[!ht]
\tabletypesize{\scriptsize}
\addtolength{\tabcolsep}{-3pt}
\tablecolumns{5}
\tablecaption{Calibration between Synthetic and Observed Flux Densities}
\tablehead{
\colhead{Band} & \colhead{$\alpha_{0}$} & \colhead{$\beta_{0}$} & \colhead{$\alpha_{1}$} & \colhead{$\beta_{1}$}
}
\startdata
IRC S7 & $0.948 \pm 0.009$ & $0.099 \pm 0.012$  & $0.988 \pm 0.007$ & $0.052 \pm 0.010$ \\
IRAC4 & $0.940 \pm 0.012$ & $0.108 \pm 0.014$  & $0.973 \pm 0.009$ & $0.071 \pm 0.011$ \\
IRC S11 & $0.971 \pm 0.006$ & $-0.003 \pm 0.008$ & $1.003 \pm 0.007$ & $-0.039 \pm 0.009$ \\
W3 & $1.035 \pm 0.007$ & $-0.017 \pm 0.008$  & $1.069 \pm 0.008$ & $-0.055 \pm 0.009$ \\
IRC L15 & $0.875 \pm 0.008$ & $0.148 \pm 0.010$  & $0.911 \pm 0.006$ & $0.107 \pm 0.008$ \\
MIPS24 & $0.983 \pm 0.014$ & $-0.020 \pm 0.018$  & $1.009 \pm 0.012$ & $-0.051 \pm 0.016$ \\
IRC L24 & $0.926 \pm 0.012$ & $0.123 \pm 0.018$  & $0.949 \pm 0.010$ & $0.099 \pm 0.015$ \\
\enddata
\tablecomments{Col. (1): Band name. Col. (2): Slope before scaling the spectrum. Col. (3): Zero point before scaling the spectrum. Col. (4): Slope after scaling the spectrum. Col. (5): Zero point after scaling the spectrum.}
\label{tab:table2}
\end{deluxetable} 

\section{PAH Decomposition} \label{section:sec3}

After binning the spectral datacube, we have 106 complete low-resolution spectra, among them five from the nuclear region, 26 from the central region, 15 from the spiral arms, 20 from the inner ISM region, and 40 from the outer ISM region. This section first introduces the methodology for measuring PAHs using complete IRS spectra, and then it assesses the feasibility of extracting PAH measurements using incomplete IRS (only SL or LL) spectra supplemented by broadband photometry.

\subsection{PAH Measurement with Complete IRS Spectra}\label{section:sec3.1}

Figure~\ref{fig:fig02} illustrates the decomposition of PAH emission from the complete, low-resolution IRS spectra on sub-kpc scales, closely following the methodology of \cite{Xie et al. 2018a}. After first removing ionic emission lines that are isolated from the main PAH features, we fit the $\sim 5-38\, \mum$ spectrum with a four-component model consisting of a theoretical PAH template plus dust continuum represented by three modified blackbodies of different temperatures, all subject to dust attenuation by foreground extinction. We calculate the theoretical PAH template assuming a starlight intensity $U\,=\,1$ times the interstellar radiation field of the ISM in the solar neighborhood (\citealt{Mathis et al. 1983}).  We adopt grain sizes $3.5\Angstrom < a < 20\Angstrom$, and we treat stochastic heating of small grains (\citealt{Draine & Li 2001}). \cite{Xie et al. 2018a} show that the PAH spectrum is relatively insensitive to radiation intensities up to $U \approx 10^4$. 

\cite{Xie et al. 2018a} demonstrate that the above template-fitting method can effectively decompose the PAH emission from the full low-resolution (SL+LL) IRS spectra of a wide range of environments, from individual Galactic high-latitude clouds to extragalactic systems of ever-increasing complexity, including normal star-forming galaxies to low-luminosity AGNs, powerful quasars, and heavily obscured infrared-luminous galaxies. An even larger and more diverse sample of star-forming galaxies has been decomposed by \cite{Xie & Ho 2019} using our technique. Furthermore, \cite{Xie et al. 2018a} compare the PAH strengths measured from their template-fitting method with those obtained using other methods in the literature, including {\tt PAHFIT} (\citealt{Smith et al. 2007b}), spline fit (\citealt{Peeters et al. 2002}), and {\tt CAFE} (\citealt{Marshall et al. 2007}). Good agreement is found with the results based on {\tt PAHFIT} and {\tt CAFE}, while the spline fit method yields systematic differences that can be understood as a consequence of the effect of the spline fit on the wings of the PAH features (see Figures 17--19 in \citealt{Xie et al. 2018a}).

We use the Bayesian Markov chain Monte Carlo (MCMC) procedure {\tt emcee} in the {\tt Python} package to determine the posterior distribution of each best-fit parameter, whose median and standard deviation of the distribution are taken as the estimate of the parameter and its error. The standard deviation of the posterior distribution is correlated with the input uncertainty of the data (\eg \citealt{Kelly 2007}). Because {\tt CUBISM} provides relatively small statistical uncertainties for the IRS mapping-mode observations, the formal errors of the measurements are very small and possibly unrealistic. We return to this point later. 

We define the total PAH emission by integrating the best-fitting PAH template over the region $5-20\, \mu$m, which captures most of the prominent PAH features within the IRS spectral range (see inset in Figure~\ref{fig:fig02}). This differs slightly from the approach of \cite{Xie & Ho 2019}, which, based on other considerations, adopts instead an integration region of $5-15\, \mu$m. However, given a constant theoretical PAH spectrum, the integrated PAH flux in the $5-20\, \mu$m region is simply a constant multiple (1.147) of that in the $5-15\, \mu$m region. We further note that we obtain the total PAH emission by integrating directly the best-fit decomposed PAH spectrum, while \cite{Xie & Ho 2019} perform their integration on the continuum-subtracted emission-line spectrum (observed spectrum minus best-fit continuum model). Our strategy is motivated by situations where we lack complete spectral coverage (Section~\ref{section:sec3.2}), and, in any case, in practice there is little actual difference between the final results of the two approaches.\footnote{ For the spaxels of M51, the difference between these two sets of PAH measurements is $-0.006 \pm0.009$ dex.}

Of course, one major disadvantage of measuring PAH based on a fixed template is that we sacrifice all information on possible spectral variations. Therefore, where the data permit, it is important to retain the flexibility of working directly with the residual PAH spectrum after subtracting the best-fit continuum components. This will be the approach to examine the relative strengths of individual PAH features after decomposing them using a series of Drude profiles (\citealt{Draine & Li 2007}), as described in \citealt{Xie et al. 2018a}. We note that the template-fitting method employed here has been optimized and fully tested only for applications to star-forming galaxies, for which a single theoretical PAH template suffices. While the current template can accommodate modest variations in the PAH spectrum across various star-forming galaxies (\citealt{Xie et al. 2018a}) and on sub-galactic scales within individual galaxies such as M51 (Section~\ref{section:sec4.2}), more extreme spectral variations may require consideration of more diverse PAH templates, which are beyond the scope of this work.

\begin{figure*}[!ht]
\center{\includegraphics[width=1\linewidth]{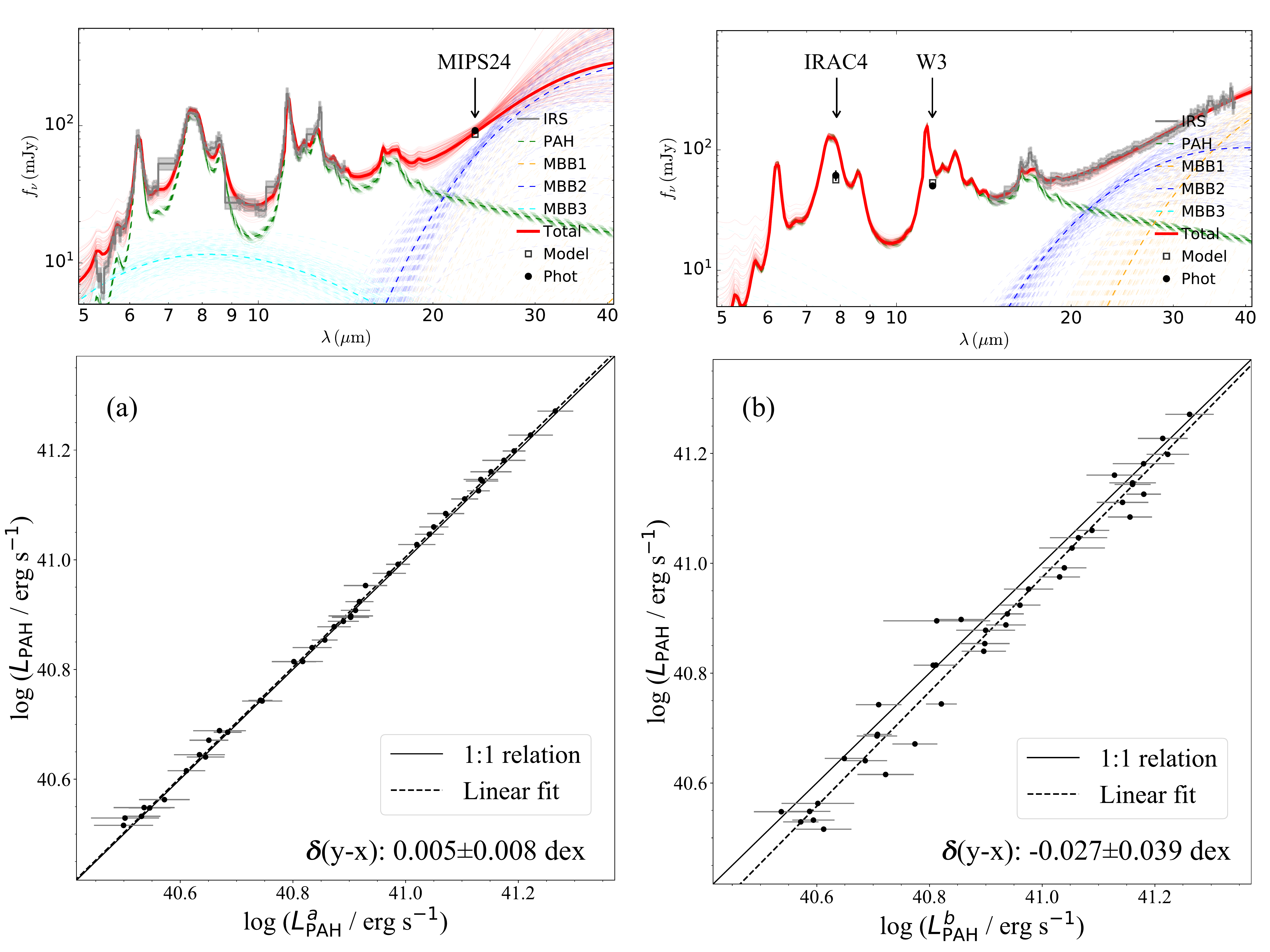}}
\caption{{PAH decomposition results for the complete IRS spectra (y-axis) and for the spectral energy distribution based on (a) SL spectra plus MIPS24 photometry and (b) LL spectra plus IRAC4 and W3 photometry. The median and standard deviation of the difference between the y-axis and x-axis are given in the lower-right corner of each panel.}
\label{fig:figC}}
\end{figure*}

\subsection{PAH Measurement Combining IRS Spectra and Photometric Data}\label{section:sec3.2}

The PAH measurement method described in the preceding subsection requires complete (SL and LL) spectral coverage over $5-38\,\mu$m. Only limited portions of nearby galaxies---usually their central regions---satisfy this condition because the configuration of IRS is such that the SL and LL slits are nearly perpendicular to each other, and the two slits also differ in aperture size\footnote{\url{https://irsa.ipac.caltech.edu/data/SPITZER/docs/irs/irsinstrumenthandbook/4/}}. These practical constraints imply that in the outer regions of extended sources often only SL or LL data are available for IRS mapping-mode observations. To maximize our ability to extract PAH measurements over the largest possible range of environments, we devise a method to decompose the MIR spectral energy distribution that complements the partial IRS spectral coverage with MIR photometry. The relevant photometric data, summarized in Table~\ref{tab:table1}, derive from Spitzer IRAC4 (7.87~$\mum$) and MIPS24 (23.68~$\mum$), WISE W3 (11.56~$\mum$), and AKARI IRC (7.0, 11.0, 15.0, 24~$\mum$) bands. Here we highlight the subset (IRAC4, W3, and MIPS24) that is most widely available for the largest number of nearby galaxies. Specifically, we discuss the two most useful combinations: (1) IRS SL spectra plus MIPS24 photometry, and (2) IRS LL spectra plus photometry from IRAC4 and W3.

We test our procedure using, as reference, spectra extracted from the central region (region 2) and spiral arms region (region 3) of M51, locations where PAH emission is prominent and background subtraction has minimal influence. As the error bars of the spectroscopic data are usually smaller than those of the photometric data, during the fitting process we artificially add 10\% additional uncertainty to the spectra to better balance the weighting of the two types of data. This value is comparable to the uncertainty of the photometric data (see also \citealt{Draine et al. 2007}). Figure~\ref{fig:figC} compares PAH measurements from the complete IRS spectra with those derived from the joint fitting of partial spectra and photometry. The combination of SL spectra plus MIPS24 photometry (Figure~\ref{fig:figC}a) effectively recovers the true PAH luminosity with no systematic bias and almost negligible scatter ($0.005\pm0.008$ dex):

\begin{align}\label{equ:equonlySL}
\begin{aligned}
&{\rm log}\,L_{\rm PAH}  = \\&(1.005\pm0.026)({\rm log}\,L_{\rm PAH}^{a}  - 40.9) + (40.905\pm0.005).
\end{aligned}
\end{align}

\noindent
Using IRAC4 and W3 to supplement LL spectra gives a slight bias and larger scatter ($0.027\pm0.039\,\rm dex$; Figure~\ref{fig:figC}b),

\begin{align}\label{equ:equonlyLL_a}
\begin{aligned}
&\log\,L_{\rm PAH} = \\&(1.041\pm0.032)(\log\,L_{\rm PAH}^{b}  - 40.9) + (40.870\pm0.007).
\end{aligned}
\end{align}

\noindent
although both are still very small and acceptable for many scientific applications, in light of the far fewer spectral features available to constrain PAH emission in the LL spectrum. 

We conclude that even if only partial IRS spectra are available, our template-fitting method can still provide good measurements of the integrated PAH luminosity, by combining the partial spectra with suitably chosen MIR photometry. This will enable us to take advantage of the more widely available single-mode (either SL or LL) IRS spectra, which have previously not been fully explored. The above calibrations (Equations~\ref{equ:equonlySL} and \ref{equ:equonlyLL_a}) enable the measurement of spatially resolved SFRs based on the integrated PAH luminosity for regions of nearby galaxies where only partial IRS mapping-mode data exist. As noted at the end of Section~\ref{section:sec3.1}, our template-fitting framework of PAH measurement is currently limited to spectra dominated by star-forming regions. The accuracy of our decomposition and our ability to estimate the total PAH emission based on incomplete spectra strongly depend on this assumption, and therefore the uncertainties of Equations~\ref{equ:equonlySL} and \ref{equ:equonlyLL_a} should be viewed as lower limits. We caution that Equations~\ref{equ:equonlySL} and \ref{equ:equonlyLL_a}, and the Xie et al. (2018a) template-fitting technique on which they are based, may not be applicable to star-forming systems of very low metallicity, regions dominated by evolved stars, or environments strongly affected by AGNs.

\section{Application to M51}\label{section:sec4}

\begin{figure*}[!ht]
\center{\includegraphics[width=1\textwidth]{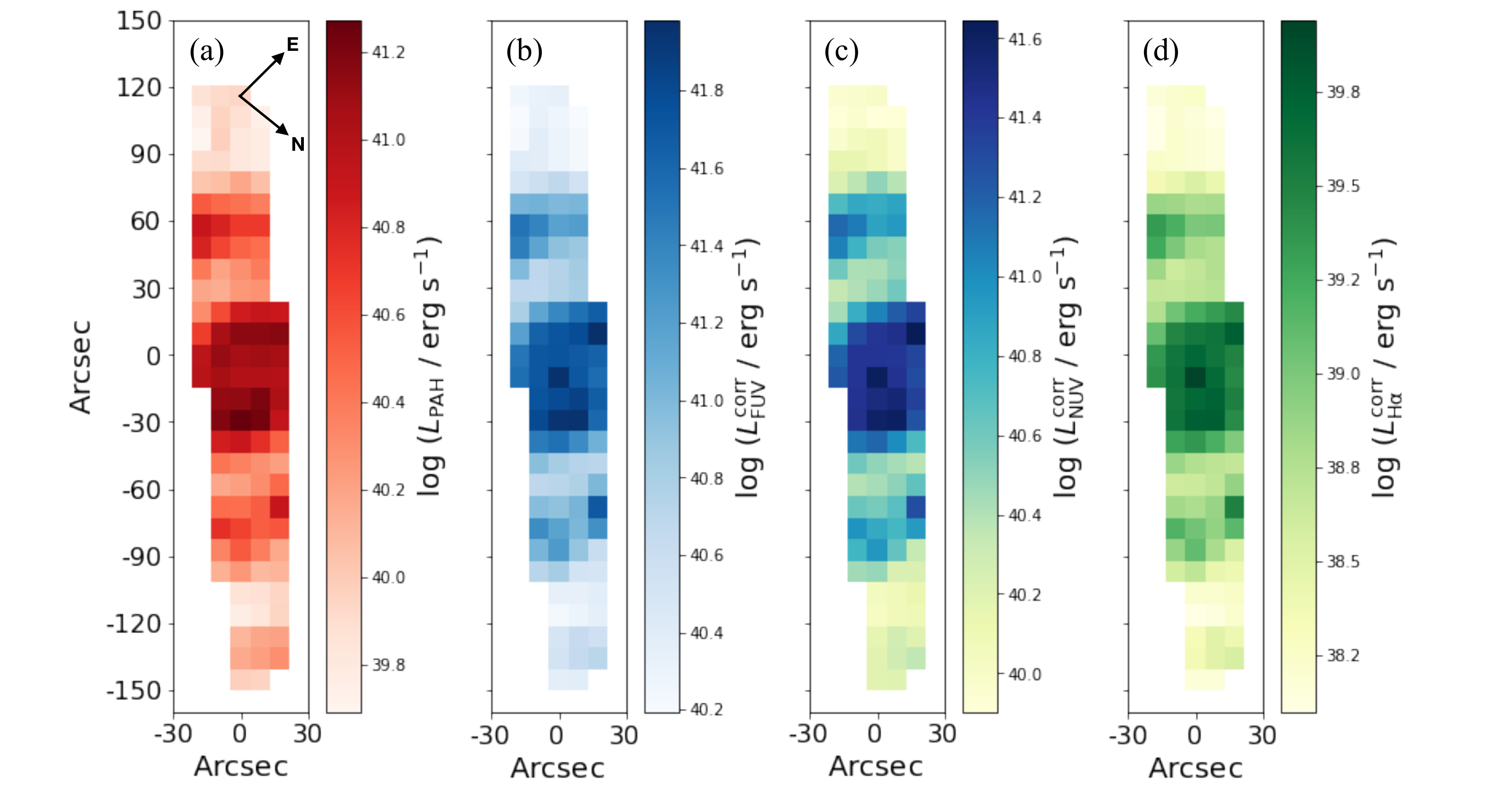}}
\caption{Spatial distribution of star formation indicators in the central $\sim 40\arcsec \times 240\arcsec$ of M51, for luminosity in (a) PAH ($5-20\,\mu$m) emission and extinction-corrected (b) FUV emission, (c) NUV emission, and (d) H$\alpha$ emission. The maps are centered on the nucleus of the galaxy, and the orientation for north and east are indicated in panel (a).
\label{fig:figmap}}
\end{figure*}

The above-described procedures enable us to derive spatially resolved (sub-kpc scale) PAH distributions for nearby galaxies with available IRS mapping-mode observations.  To showcase the scientific potential of these data, here we highlight a few simple, illustrative applications to M51, whose central $\sim 40\arcsec \times 240\arcsec$ (1.6 kpc $\times$ 9.5 kpc) region has complete IRS observations in both the SL and LL modes.  Figure~\ref{fig:figmap} shows the PAH ($5-20\,\mu$m) emission-line map, alongside separate views of the galaxy as imaged through tracers of star formation in FUV, NUV, and H$\alpha$ emission.

\begin{figure*}[!ht]
\center{\includegraphics[width=1\textwidth]{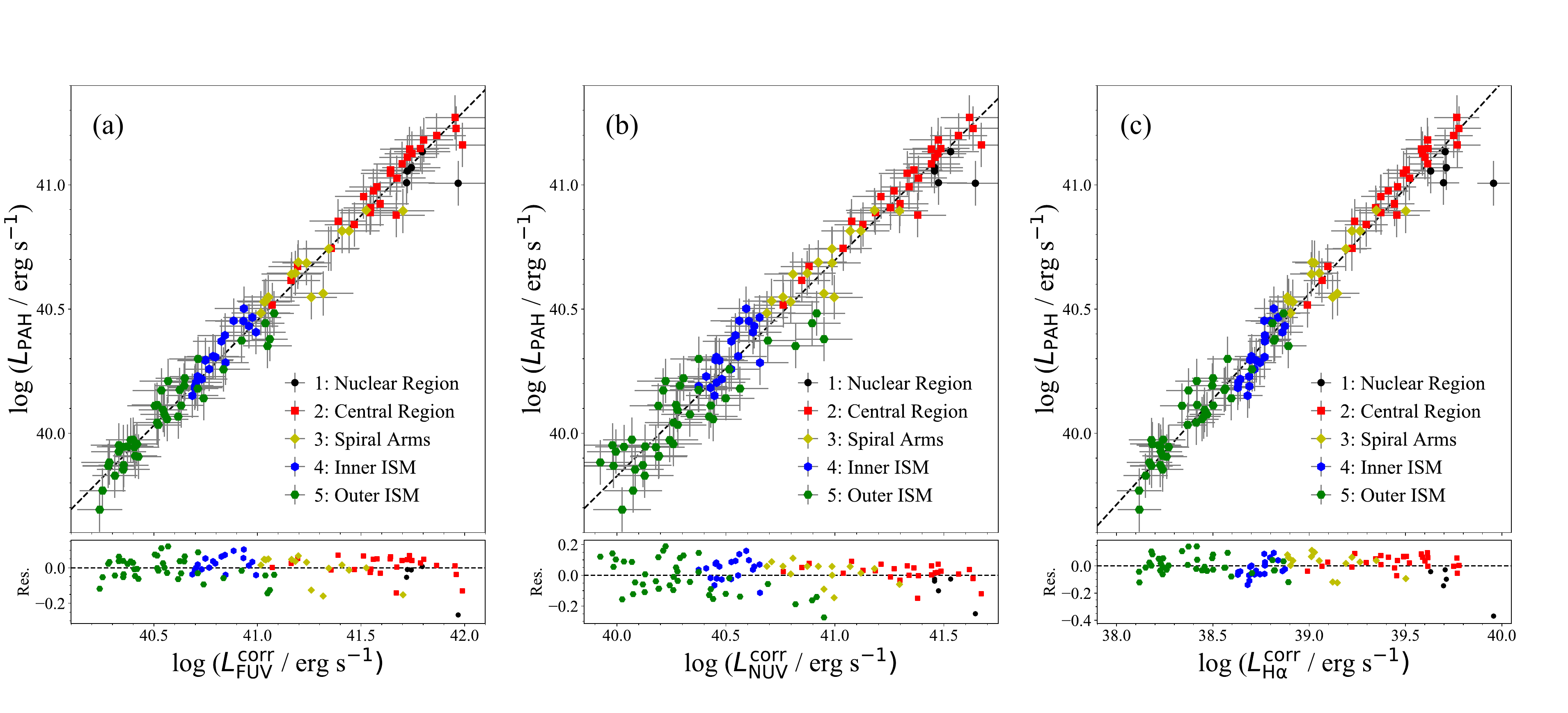}}
\caption{{Comparison of PAH ($5-20\,\mu$m) luminosity with extinction-corrected (a) FUV, (b) NUV, and (c) H$\alpha$ luminosity. Data points with different colors and symbols correspond to the five regions of M51, as indicated in Figure~\ref{fig:fig01}. The subpanels show the residuals between the best fit (dashed line) and the observations. }
\label{fig:fig03}}
\end{figure*}

\subsection{Spatially Resolved Correlation between PAH Emission and other Star Formation Indicators}\label{section:sec4.1}

Figure~\ref{fig:fig03} shows the correlations between PAH luminosity $L_{\mathrm{PAH}}$ and monochromatic luminosities in the two UV bands, defined as $L_{\rm band} =\nu L_{\nu}({\rm band})$, and luminosity in H$\alpha$. The error budget on PAH luminosity includes several components combined in quadrature, including  the uncertainty associated with the PAH decomposition (\citealt{Xie et al. 2018a}), a 5\% uncertainty on the absolute flux calibration for IRS (\citealt{Houck et al. 2004}), and the uncertainty incurred from the process of matching the spectral and photometric data (Section~\ref{section:sec2.2.5}), which we conservatively estimate to be 20\% (Figure~\ref{fig:figB}). The error budget for the photometry-based luminosities includes uncertainties from Poisson noise from the signal, background subtraction, variations in the background, and absolute flux calibration.

We correct the UV bands and H$\alpha$ for Galactic extinction, assuming $E(B - V) = 0.03$ mag (\citealt{Schlafly & Finkbeiner 2011}) and the Galactic extinction curve of \cite{Cardelli et al. 1989} for $R_V = 3.1$. For the UV bands, we assume, as do  \cite{Gil de Paz et al. 2007}, $A_{\rm FUV} = 7.9E(B - V)$ and $A_{\rm NUV} = 8.0E(B - V)$, and we correct for internal extinction using MIPS24 photometry and the prescriptions $L_{\rm FUV}^{\rm corr} = L_{\rm FUV}^{\rm obs} + 6.0\,L_{\rm 24\mum}$ (\citealt{Liu et al. 2011}) and $L_{\rm NUV}^{\rm corr} = L_{\rm NUV}^{\rm obs} + 2.26\,L_{\rm 24\mum}$ (\citealt{Hao et al. 2011}), where $L_{\rm FUV}^{\rm obs}$ and $L_{\rm NUV}^{\rm obs}$ represent quantities corrected for Galactic extinction. Differences in the assumptions used by \cite{Hao et al. 2011} and \cite{Liu et al. 2011} to estimate the internal extinction introduce an uncertainty at the level of $\sim 30\%$, which is added to the final error budget for the UV luminosities.

In the case of H$\alpha$, we adopt $A_{\rm H\alpha} = 2.535 E(B-V)$ (\citealt{Kennicutt et al. 2007}). Correction for internal extinction again uses MIPS24 photometry, following the prescription $L_{\rm H\alpha}^{\rm corr} = L_{\rm H\alpha}^{\rm obs} + 0.031\,L_{\rm 24\mum}$ of \cite{Calzetti et al. 2007}, with $L_{\rm H\alpha}^{\rm obs}$ the Galactic extinction-corrected $\rm H\alpha$ luminosity. The coefficient for the correction term (0.031), derived from the observations of spatially resolved H~{\small II} regions in galaxies drawn from the SINGS sample, is larger than that (0.020) of \cite{Kennicutt et al. 2009}, which was determined from global measurements of galaxies. These two choices affect the extinction-corrected H$\alpha$ luminosity at the level of $\sim20\%$, which we include into the final error budget.

We observe tight correlations between the PAH luminosity and the luminosity in FUV, NUV, and H$\alpha$ across M51, from the galaxy nucleus, to the central star-forming region, spiral arms, and areas in between the spiral arms. Linear regression analysis with the {\tt Python} package {\tt linmix} (\citealt{Kelly 2007}) yields

\begin{equation}
\begin{aligned}\label{equ:calFUV}
&{\rm log}\,L_{\rm PAH} = \\&(0.842\pm0.027)({\rm log}\, L_{\rm FUV}^{\rm corr} - 41.0) + (40.455\pm0.013),
\end{aligned}
\end{equation}

\begin{align}
\begin{aligned}\label{equ:calNUV}
&{\rm log}\,L_{\rm PAH} = \\&(0.865\pm0.029)({\rm log}\, L_{\rm NUV}^{\rm corr} - 41.0) + (40.696\pm0.016),
\end{aligned}
\end{align}

\noindent
and

\begin{align}\label{equ:calHa}
\begin{aligned}
&{\rm log}\,L_{\rm PAH} = \\&(0.849\pm0.026)({\rm log}\, L_{\rm H\alpha}^{\rm corr} - 39.0) + (40.561\pm0.013).
\end{aligned}
\end{align}

\noindent
The observed scatter of the above relations is very small, only $\rm 0.064\, dex$, $\rm 0.091\, dex$, and $\rm 0.073\, dex$, respectively; the corresponding intrinsic scatter is $\rm 0.022\, dex$, $\rm 0.027\, dex$, and $\rm 0.022\, dex$. Thus, to the extent that the FUV and NUV bands and H$\alpha$ emission faithfully trace young stars, PAH emission also traces regions of star formation on sub-galactic scales in M51. 

\subsection{Variation of PAH Emission with Environment}\label{section:sec4.2}

PAHs are a ubiquitous constituent of the interstellar medium, displaying a spectrum that, to first order, changes surprisingly little across a wide range of galaxy environments (\citealt{Xie et al. 2018a}). A major motivation of this and future work in this series is to use spatially resolved spectroscopy to examine the extent to which this holds on sub-galactic scales over a broad range of physical conditions. We further aim to study the validity of PAH as a SFR estimator by comparing it directly to independent indicators of star formation activity, again in a spatially resolved manner. In this first pilot project, we succeed in resolving the central $\sim 40\arcsec \times 240\arcsec$ of M51 into 106 spatially independent spaxels, spanning environments as diverse as the nuclear region that hosts a low-luminosity AGN to diffuse areas occupying inter-arm regions and the outer extent of the disk.

\begin{figure}[!ht]
\center{\includegraphics[width=1\linewidth]{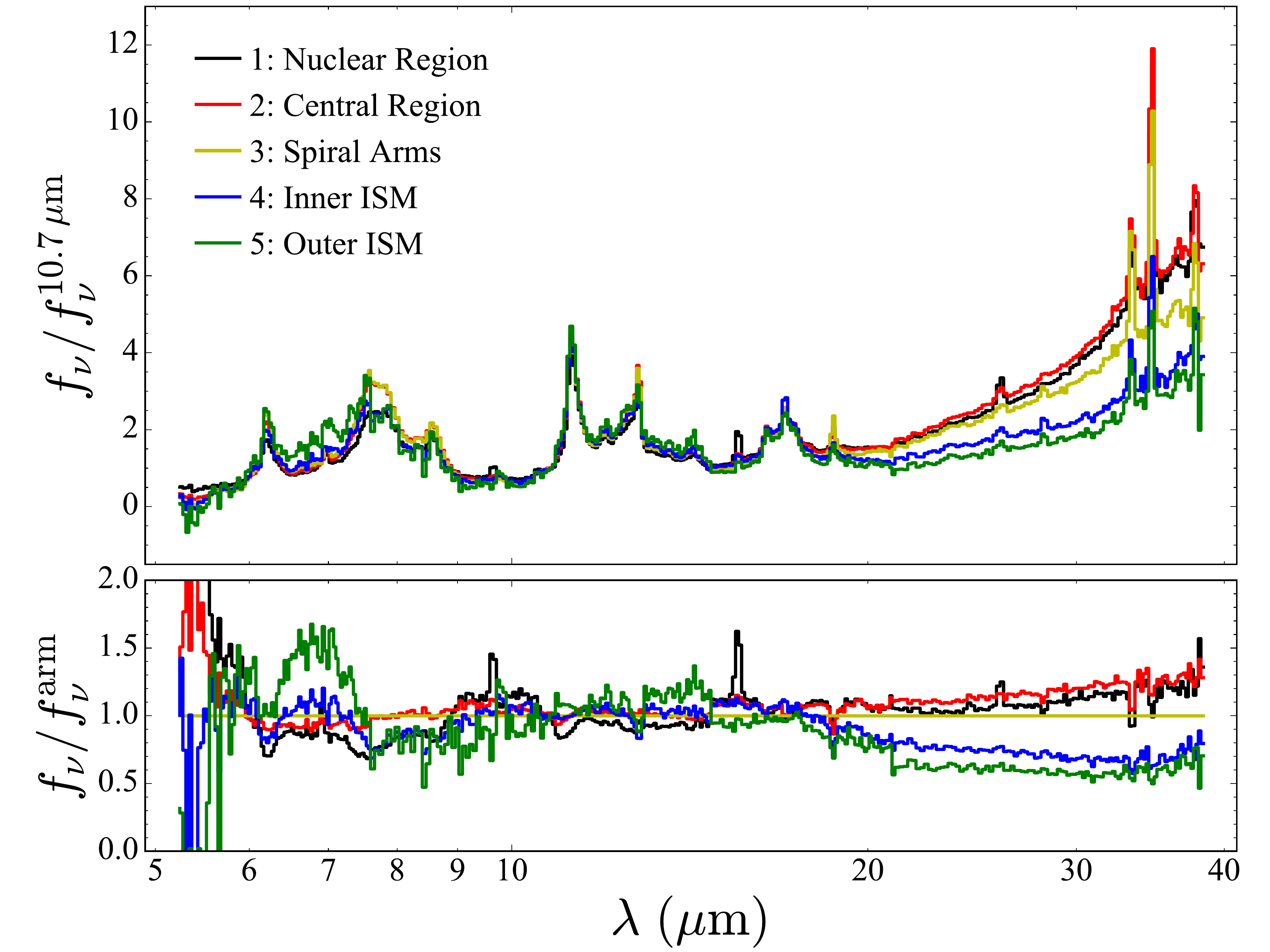}}
\caption{{(Top) Integrated IRS spectra extracted from the five different regions, normalized with the flux density at 10.7~$\mu$m. (Bottom) Ratio of the spectra in the top panel relative to the spectrum of the spiral arm region.}
\label{fig:fig06}}
\end{figure}

To illustrate possible higher order variations of the PAH spectrum in different environments, Figure~\ref{fig:fig06} shows the integrated spectra of the spaxels extracted from the five different regions defined in Figure~\ref{fig:fig01}. The overall spectral shape from 5 to $20\,\mum$ is remarkably uniform across the various regions, reinforcing not only the general homogeneity of the PAH spectrum (\citealt{Xie et al. 2018a}) but the widespread presence of hot (300--400 K;  see modified blackbody plotted in cyan in the top-right inset of Figure~\ref{fig:fig02}) dust continuum emission that \cite{Xie et al. 2018b} attribute to single-photon heating of nanometer-sized grains. By contrast, greater diversity appears in the continuum longward of $\sim 20\,\mum$, where cold dust dominates the radiation. The two diffuse ISM regions (regions 4 and 5), in particular, depart conspicuously from the rest by their flatter continuum slopes.

\begin{figure}[!ht]
\center{\includegraphics[width=1\linewidth]{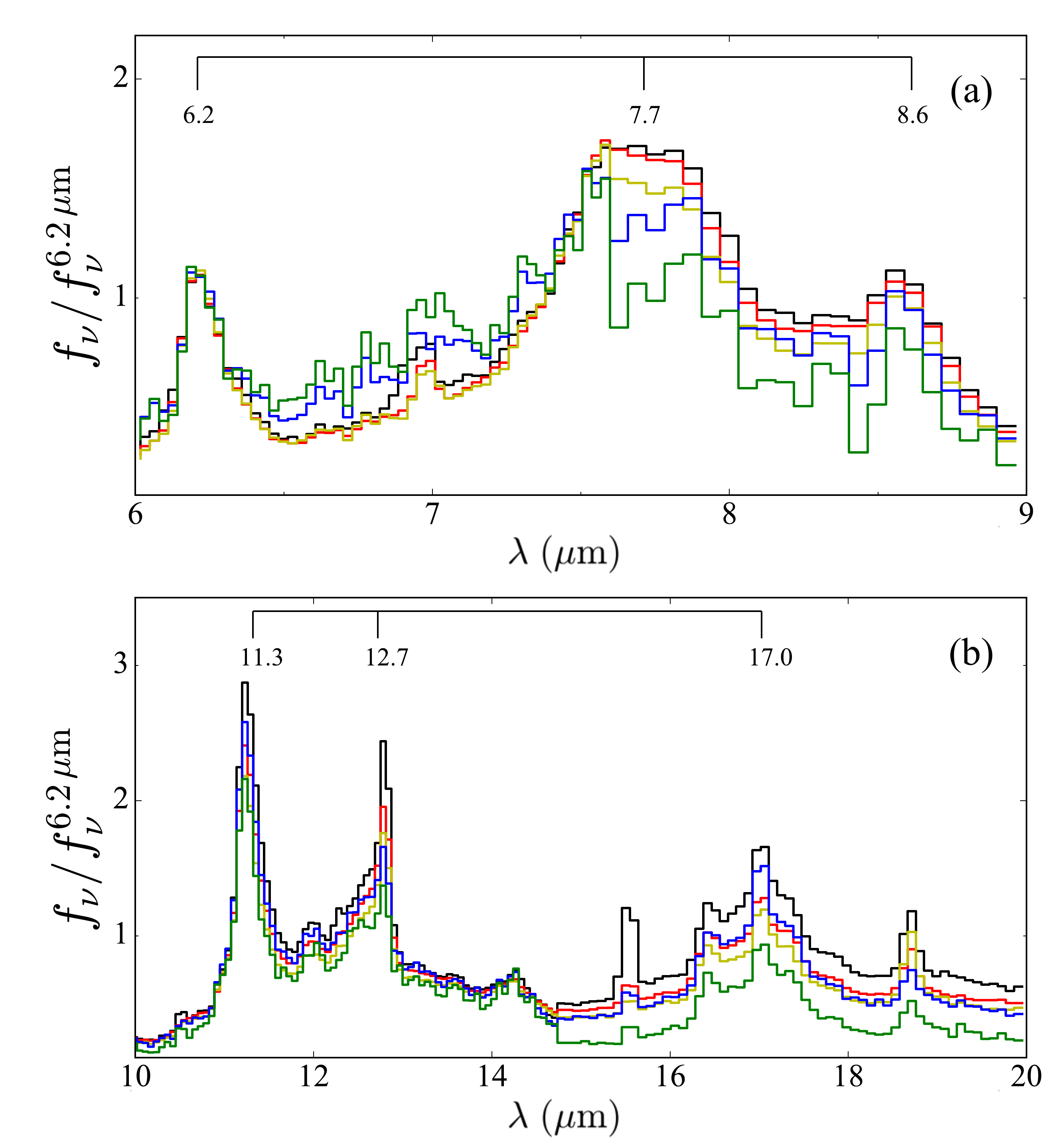}}
\caption{Zoom in of the spectra shown in Figure~\ref{fig:fig06}, after continuum subtraction. Sample spectra extracted from the five different regions, renormalized to the peak strength of the 6.2~$\mu$m feature, for the wavelength range (a) $6-9\ \mu$m and (b) $10-20\ \mu$m. The main PAH features are labelled.
\label{fig:fig06z}}
\end{figure}

Figure~\ref{fig:fig06z} expands the $5-20\,\mum$ region to facilitate a clearer view of the main PAH features, after subtracting the best-fitting continuum model and renormalizing the spectra to the peak of the 6.2~$\mu$m feature to highlight the relative variation of different PAH features. Despite the overall similarity among the various regions, some systematic differences can be discerned. For instance, when normalized relative to the 6.2~$\mu$m feature, all the PAH features at longer wavelengths are stronger for the nuclear region, which suggests that the 6.2~$\mu$m feature, mostly radiated by small, ionized PAHs, is weaker in the nuclear (central $\sim$400 pc) region, where an AGN resides, than elsewhere. This may indicate the preferential destruction of small PAHs in the vicinity of the nucleus (e.g., \citealt{Smith et al. 2007b}; \citealt{Diamond-Stanic & Rieke 2010}). Indeed, the harsh radiation field of the AGN in M51 may influence the total PAH output, as indicated by the possible systematic suppression of the integrated PAH luminosity relative to the UV and H$\alpha$ for the spaxels in the nuclear region (black points in Figure~\ref{fig:fig03}), especially for the centralmost spaxel. Meanwhile, the inner $\sim$2.5 kpc and the spiral arms, both sites of ongoing star formation, display stronger 7.7 and 8.6~$\mu$m features relative to 6.2~$\mu$m feature. The 11.3~$\mum$ feature is constant in star-forming regions but varies in other parts of the galaxy. Note that the spectrum of the outermost portions of M51's disk, sampled by region 5, exhibits a broad excess between $\sim 6.5$ and 7.5~$\mu$m. We have verified that this feature is a calibration artifact affecting regions of low surface brightness observed in the SL2 order.\footnote{This effect is also described in the IRS instrument manual: {\url{https://irsa.ipac.caltech.edu/data/SPITZER/docs/irs/irsinstrumenthandbook.}}}

The degree of coupling between PAH and UV emission may depend on the environment within the galaxy. This is especially noticeable in the NUV band (Figure~\ref{fig:fig03}b), where the spaxels from the central star-forming region and spiral arms correlate moderately more tightly (scatter = $\rm 0.059\, dex$) than the two ISM regions (scatter = $\rm 0.067\, dex$). Does this indicate that in the inter-arm regions PAHs are not heated by UV photons from young stars but instead by non-ionizing radiation from evolved stars (e.g., \citealt{Hoopes & Walterbos 2000}; \citealt{Hoopes et al. 2001}; \citealt{Calzetti et al. 2005})?  Interestingly, the correlation between PAH and UV luminosity has a slope that is slightly, but significantly, below unity (the slope is 0.84 for FUV and 0.87 for NUV), in qualitative agreement with previous studies (\citealt{Calzetti et al. 2005}; \citealt{Wu et al. 2005}). We will explore these and other related issues in future work.

\section{Discussion and Summary}\label{section:sec5}

We present a new method to extract spatially resolved MIR spectra of nearby galaxies, using Spitzer IRS mapping-mode observations that simultaneously preserve the information content on galaxy properties while maximizing spatial resolution and spatial coverage. Our primary goal is to investigate MIR spectral diagnostics on sub-kpc scales, with special emphasis on star formation indicators such as PAH emission. Compared to previous adaptive binning methods for integral-field spectroscopy, such as Voronoi tessellations (\citealt{Cappellari & Copin 2003}), our new optimal binning method considers the spatial distribution of galaxy properties by assessing the redundancy of spectra. We perform simulations to determine the minimum S/N threshold needed per spaxel to ensure robust measurement of the total $5-20\,\mu$m PAH emission, using the PAH template-fitting method of \cite{Xie et al. 2018a}. Our method applies to IRS mapping-mode observations having low-resolution spectral coverage over the full $\sim 5-38\,\mu$m region, or to observations having only partial spectral coverage but that can be supplemented by suitably chosen MIR photometric imaging. 

As a pilot test case, we apply our method to archival IRS observations of the central $\sim 40\arcsec \times 240\arcsec$ (1.6 kpc $\times$ 9.5 kpc) region of the nearby spiral galaxy M51. The mapped area is sampled by 106 independent 0.4 kpc spaxels, which are carefully matched to GALEX UV and ground-based H$\alpha$ images. We show that the integrated ($5-20\,\mu$m) PAH luminosity of the individual spaxels tightly correlates with the extinction-corrected luminosity in FUV, NUV, and H$\alpha$ with scatter $\lesssim 0.1$ dex, over regions as diverse as the galaxy nucleus, the circumnuclear star-forming region, spiral arms, and inter-arm regions. These tight relations strongly support the notion that PAHs serve as an effective SFR indicator, although the slight departure from linearity in the relations suggests that PAHs may be partly heated by non-ionizing radiation from evolved stars, and PAHs may be partly destroyed or modified near the active nucleus.

Despite the overall gross similarity of the PAH spectrum across M51, some subtle regional differences can be clearly discerned. The 6.2~$\mum$ feature is suppressed in the vicinity of the active nucleus, while the 7.7 and 8.6~$\mum$ features are enhanced in the central star-forming region and spiral arms. There is a dispersion of 11.3~$\mum$ feature relative to 6.2~$\mum$ feature among different regions. Forthcoming work will examine these and other trends using a larger sample of galaxies spanning a range of physical properties.

\acknowledgments
This work is supported by the National Science Foundation of China (11721303, 11991052) and the National Key R\&D Program of China (2016YFA0400702). Y.X. is supported by the National Natural Science Foundation of China for Youth Scientist Project (11803001). We thank the anonymous referee for helpful comments and suggestions to improve the presentation of our paper. We are grateful to V. Lebouteiller for helpful discussions about the convolution of the IRS datacube and F. Egusa for kindly providing the PSFs of the four AKARI/IRC images used in this paper. This research made use of {\tt Montage}, funded by the National Aeronautics and Space Administration's Earth Science Technology Office, Computational Technologies Project, under Cooperative Agreement Number NCC5-626 between NASA and the California Institute of Technology. The code is maintained by the NASA/IPAC Infrared Science Archive. 

\appendix
\renewcommand\thefigure{\thesection}
\section{S/N Criterion for Binning the Spectral Datacube} \label{sec:appexA}

\begin{figure}[!ht]
\center{\includegraphics[width=0.65\linewidth]{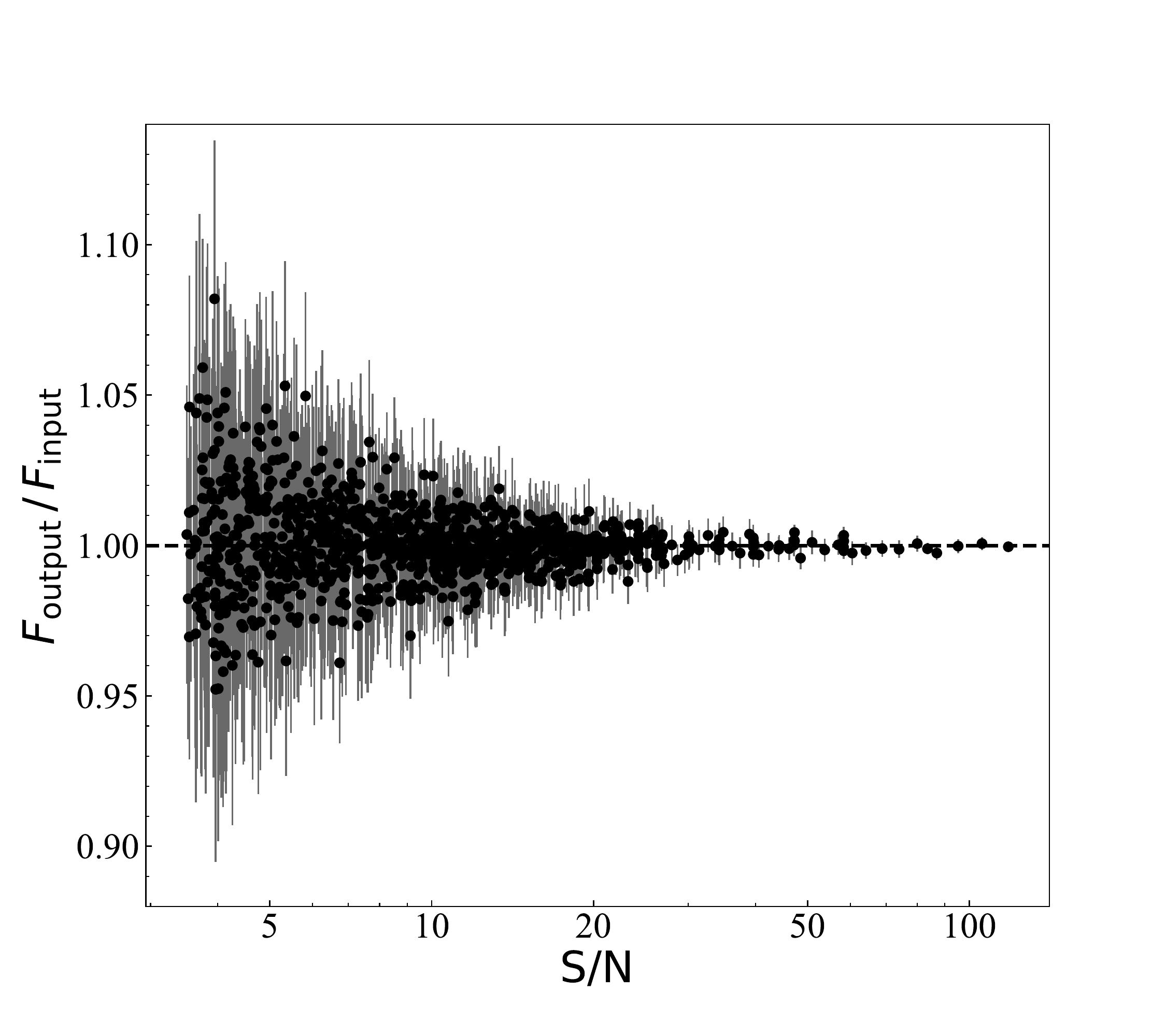}}
\caption{The effect of S/N on the ratio of the output and input integrated $5-20\,\mu$m PAH flux. These simulations serve as a guide to determine the minimum S/N threshold for the optimal bin size of the spaxels (see Section~\ref{section:sec2.2.4}).
\label{fig:figA}}
\end{figure}

We perform mock simulations to determine the S/N criterion for optimally binning the pixels into final spaxels (Section~\ref{section:sec2.2.4}). Starting with all the spaxels in the datacube that have complete (5--38~$\mum$) IRS spectral coverage, we first combine them to generate a spectrum with the maximum S/N. We then decompose the high-quality spectrum with the technique described in Section~\ref{section:sec3.1} to derive the integrated (5--20~$\mum$) PAH flux, which serves as the input flux ($F_{\rm input}$) of the simulations. Then, we add Gaussian noise to the input spectrum to generate a series of $\sim\,1400$ noisier mock spectra with S/N ranging from $\sim 3$ to 100, and then remeasure the output PAH flux ($F_{\rm output}$). Figure~\ref{fig:figA} illustrates the variation of $F_{\rm output}\,/\,F_{\rm input}$ as a function of S/N. As expected, our ability to recover the input PAH flux decreases with decreasing S/N. The uncertainty of $F_{\rm output}\,/\,F_{\rm input} \approx 1\%$ at S/N = 30 and increases to $\sim 10\%$ at S/N = 5. 


\end{document}